# Efficient Calculation of Stabilization Parameters in RF Power Amplifiers

Libe Mori, Ibone Lizarraga, Aitziber Anakabe, Juan-Mari Collantes, Vincent Armangaud, Geoffroy Soubercaze-Pun

*Abstract*— This paper proposes an efficient method for the calculation of the stabilization parameters in RF power amplifiers operating in periodic large-signal regimes. Stabilization is achieved by applying the principles of linear control theory for Periodic Linear Time-Varying (PLTV) systems. A numerical method is proposed to obtain the Harmonic Transfer Function that represents the system linearized around the large-signal steady state. Then, a feedback analysis is performed to calculate the closed-loop poles of the PLTV system. The proposed approach is demonstrated with two examples. Firstly, a three-stage amplifier that exhibits a low-frequency oscillation for increasing values of input power is correctly stabilized. Next, the stabilization of an unstable design that exhibits an odd-mode parametric oscillation is presented. The results of the proposed technique are compared to those obtained with the conventional parametric stability simulation. These examples serve to illustrate the capability and efficiency of the proposed approach.

## I. INTRODUCTION

DUE to their inherent nonlinear behavior, RF power amplifiers have a tendency to generate spurious signals of autonomous nature that show up depending on the power level and frequency of the input signal drive. Methods and techniques to predict this kind of unstable behavior have been amply reported in the literature [1]-[6].

Pole-zero identification is one of the existing methods for stability analysis of small-signal and large-signal regimes [6]-[8]. In this method, a frequency response representing the linearized system is approximated by a ratio of two polynomials. Commercially available tools [9] apply automatic algorithms to identify the poles and zeros of frequency responses of unknown order in limited frequency bands. The local fitting facilitates the approximation of the delays introduced by the transmission lines of microwave circuits with sufficient accuracy. These tools can be used in combination with common microwave circuit simulators [10], [11].

Once a possible oscillation is detected, the designer needs to implement effective stabilization solutions with the smallest possible impact on circuit performances. In amplifiers with multiple non-linear elements, finding the appropriate value and location for the stabilization elements is generally a non-systematic trial and error procedure. This usually involves a considerable number of parametric non-linear simulations that can be particularly time-consuming in the case of periodic large-signal regimes (harmonic–balance simulations with conversion matrix algorithm while sweeping the stabilization parameter).

In [12], a systematic technique was proposed to determine the value and location of the stabilization elements of a microwave amplifier suffering from instability at dc or small-signal regime. Given a Single-Input Single-Output (SISO) transfer function obtained at a circuit's observation port, pole placement techniques such as root-locus tracing were used to directly provide the position of the circuit poles versus the stabilization parameter without the need of further simulations.

However, the approach in [12] is not directly applicable to instabilities that are function of the power or frequency of the drive signal. Stability analysis in the presence of a periodic large-signal input requires the linearization of the large-signal steady state. Due to the side-band conversions, the result of this linearization is a Periodic Linear Time-Varying (PLTV) system that cannot be represented by a SISO transfer function. Instead, the side-band conversions inherent to the PLTV system have to be represented by a matrix of transfer functions, as in a Multiple-Input Multiple-Output (MIMO) system. Unfortunately, the root locus tracing used in [12] is only applicable to SISO systems, not to MIMO systems. MIMO approaches have been used for multi-port stabilization of large-signal steady states in previous works [13], [14]. In the context of pole-zero identification, a multi-port strategy to minimize the pole-zero cancellations (or quasi-cancellations) due to loss of

Manuscript submitted March 11, 2020. This work was supported in part by the Spanish Administration under Project PID2019-104820RB-I00 and in part by the Basque Country Government under Project IT1104-16.

L. Mori is with Grupo de Teoría de la Señal y Comunicaciones, Mondragon Unibertsitatea, Loramendi 4, 20500 Mondragon, Spain (email: libemori@gmail.com).

I. Lizarraga, A. Anakabe and J.M. Collantes are with Departamento de Electricidad y Electrónica, University of the Basque Country UPV/EHU, 48940 Leioa, Spain (e-mail: ibone.lizarraga@ehu.eus; aitziber.anakabe@ehu.eus; juanmari.collantes@ehu.eus).

V. Armengaud and G. Soubercaze-Pun are with the Centre National d'Etudes Spatiales (CNES), Toulouse Space Centre, 31401 Toulouse, France (e-mail: vincent.armangaud@cnes.fr; geoffroy.soubercaze-pun@cnes.fr).

controllability or observability and to detect suitable locations for stabilization has also been presented in [15]. In [15], "multi-port" referred to different physical observation ports, not to the virtual ports corresponding to the frequency sidebands.

There is an additional reason why application of pole-placement techniques to the PLTV system resulting from linearizing a large-signal steady state is more delicate than its application to the linearization of a dc steady state. A necessary condition for [12] to be applicable is that the steady state under analysis (a dc regime fixed by the bias conditions in that case) is not modified by the variation of the stabilization parameter. This is easy to satisfy in [12], since the technique only concerns dc regimes fixed by the bias conditions. We only need to guarantee that the stabilization parameter (a resistor, capacitor or inductance) does not modify the bias point. In general, the non-perturbation of the steady state is more difficult to ensure in periodic large-signal regimes. However, there are a number of conditions for which stabilization parameters have little or no effect on the large-signal steady state under study. One case corresponds to the inter-branch resistors introduced to stabilize odd-mode instabilities in amplifiers with power-combined topologies. These resistors are not seen by the even mode steady state. A second situation concerns the low-frequency instabilities associated to a gain expansion versus input power, which are also common in power amplifiers biased in AB and B classes. In this case, stabilization is often achieved through resistive elements inserted in the bias circuitry with minimum or no effect at the operating frequency.

In this work, the practical extension of [12] to the stabilization of large-signal periodic regimes is proposed. Given a large-signal steady state, our goal is to predict the evolution of the critical poles on the complex plane as a stabilization parameter is varied without the need of performing further non-linear simulations. The formalism of the Harmonic Transfer Function (HTF) is adopted in order to model and solve a feedback PLTV system. The *HTF(s)* is an analytical conversion matrix that models the relations between the sidebands of an input/output representation in a PLTV system. A numerical approach is proposed here to obtain the *HTF(s)* matrix by identifying input/output frequency-domain data obtained with large-signal/small-signal harmonic balance simulations. The *HTF(s)* representing the PLTV system is then fed back through the stabilization network. Solving the resulting closed-loop system, the pole evolution versus the stabilization parameter can be obtained.

The paper is organized as follows. Section II recalls the formalism of the HTF and explains the proposed method for its identification from frequency-domain data. Section III presents the feedback system that is solved to obtain the pole evolution versus the stabilization parameter. Two application examples are detailed in section IV. Run times corresponding to the proposed approach are compared to those from conventional parametric non-linear simulations. A discussion on the applicability of the approach is given in Section V.

II. HARMONIC TRANSFER FUNCTION

In this section we present the formalism of the Harmonic Transfer Function and its calculation from frequency domain circuit simulations. The HTF formalism, introduced by Wereley and Hall in [16], is a general formulation that serves to analytically relate the sidebands of an input/output representation of a PLTV system. The *HTF(s)* has the form of a conversion matrix and has widely proven its applicability for studying periodic systems of all types, from helicopter rotors to wind turbines [17]-[19].

Conversion matrix is also the basic principle behind the local stability analysis of large-signal regimes in microwave circuits [1]-[4], [20]. Reduced-order conversion matrix approaches are used to analyze the stability of power amplifiers under output mismatch effects [21] or to analyze and design frequency dividers [13]. Likewise, large-signal/small-signal harmonic balance will be used here to obtain the input/output frequency-domain data serving as starting point for the calculation of the *HTF(s)* matrix. Although conversion matrix representations are well known in microwave circuit analysis, we review here the formalism of the HTF for completeness.

*A. General Formalism of the HTF*

In [16] a matrix of transfer functions for mapping the signal transfers of PLTV systems, called the Harmonic Transfer Function is introduced. This approach considers the signal content in each sideband as a separate and independent input. Thus, a SISO PLTV system can be modeled as a MIMO Linear Time Invariant (LTI) system.

Let us consider a nonlinear system described by the state-space equation in (1).

$$\dot{\bar{x}} = f(\bar{x}) \quad (1)$$

The $\bar{x}$ variable in (1) is the state vector and *f* is a continuous and infinitely derivable function. Let us denote $\bar{x}_0(t)$ a periodic solution of (1), with period $T=1/f_0$, and $\bar{\xi}(t)$ a small-signal perturbation around $\bar{x}_0(t)$. The linearization of the system dynamics around the periodic solution is given by the PLTV system in (2):

$$\dot{\bar{\xi}}(t) = G(t)\bar{\xi}(t) \quad (2)$$

where $G(t) = Jf(\bar{x}_0(t))$ is the Jacobian matrix evaluated along the periodic solution $\bar{x}_0(t)$.

Without loss of generality, to obtain an input/output representation of the linearized system, an scalar input signal $u(t)$ is introduced in (2) and the scalar system output $y(t)$ is defined as a linear combination of the state variables (3):

$$\begin{aligned}\dot{\bar{\xi}}(t) &= G(t)\bar{\xi}(t) + B(t)u(t) \\ y(t) &= C(t)\bar{\xi}(t) + D(t)u(t)\end{aligned} \quad (3)$$

where $G(t)$, $B(t)$, $C(t)$ and $D(t)$ are periodic matrices of appropriate dimensions, with the same period *T* as the solution $\bar{x}_0(t)$.

To obtain the *HTF(s)*, an Exponentially Modulated Periodic (EMP) input signal is applied to the system (3). Then, the $u(t)$, $y(t)$ and $\bar{\xi}(t)$ signals have the form of EMP signals. Therefore, they can be expanded as a Fourier series of a periodic signal with a fundamental frequency $f_0$, modulated by a complex exponential signal [22]:

$$u(t) = \sum_{k \in Z} U_k e^{s_k t}$$
$$y(t) = \sum_{k \in Z} Y_k e^{s_k t} \qquad (4)$$
$$\bar{\xi}(t) = \sum_{k \in Z} \bar{\xi}_k e^{s_k t}$$

where $t \geq 0$, $s_k = s + jk\omega_0$ and $s \in C$.

Using the Fourier coefficients of $G(t)$, $B(t)$, $C(t)$ and $D(t)$, the system of equations in (3) can be expanded in Fourier series as follows:

$$\sum_k s_k \bar{\xi}_k e^{s_k t} = \sum_k \left( \sum_m G_{k-m} \bar{\xi}_m + \sum_m B_{k-m} U_m \right) e^{s_k t}$$
$$\sum_k Y_k e^{s_k t} = \sum_k \left( \sum_m C_{k-m} \bar{\xi}_m + \sum_m D_{k-m} U_m \right) e^{s_k t}. \qquad (5)$$

The principle of harmonic balance implies that the equations in (5) must be fulfilled for each frequency $s_k$. Therefore, (5) can be rewritten compactly by matching the Fourier coefficients:

$$s\bar{\xi} = (\hat{G} - N)\bar{\xi} + \hat{B}U$$
$$Y = \hat{C}\bar{\xi} + \hat{D}U \qquad (6)$$

where the $\bar{\xi}$, $U$ and $Y$ are infinite vectors of Fourier coefficients; $\hat{G}$, $\hat{B}$, $\hat{C}$ and $\hat{D}$ are Toeplitz matrices of the Fourier coefficients of $G(t)$, $B(t)$, $C(t)$, $D(t)$; and $N$ is the block diagonal matrix defined in (7).

$$N = blkdiag\{jk\omega_0 I\} \qquad (7)$$

The *HTF(s)* matrix is defined as the matrix that relates the components of the output signal $Y$ to the components of the input signal $U$ for the PLTV system in (6) as follows,

$$HTF(s) = \hat{C}\left(sI - (\hat{G} - N)\right)^{-1} \hat{B} + \hat{D}. \qquad (8)$$

The *HTF(s)* takes the following form:

$$HTF(s) = \begin{bmatrix} \ddots & \vdots & \vdots & \vdots & \cdot \\ \cdots & H_0(s-j\omega_0) & H_{-1}(s) & H_{-2}(s+j\omega_0) & \cdots \\ \cdots & H_1(s-j\omega_0) & H_0(s) & H_{-1}(s+j\omega_0) & \cdots \\ \cdots & H_2(s-j\omega_0) & H_1(s) & H_0(s+j\omega_0) & \cdots \\ \cdot & \vdots & \vdots & \vdots & \ddots \end{bmatrix}. \qquad (9)$$

In principle, the *HTF(s)* matrix in (9) is an infinite dimensional matrix [16]. However, in practice, the matrix can be truncated to a finite number of significant sidebands, as long as the effects of the truncation can be considered negligible [23].

### B. Calculation of the HTF(s) matrix from Frequency-Domain Identifications

Obviously, the direct calculation of the analytical *HTF(s)* matrix [24] is not a practical option for complex microwave circuits. In this section, a numerical approach for the obtaining of the *HTF(s)* matrix, applicable to microwave circuits, will be described. The methodology constructs the *HTF(s)* matrix applying frequency-domain identification to an ensemble of input/output frequency responses obtained through conventional large-signal/small-signal harmonic balance simulations.

Given that the HTF considers an input/output representation of the system, the first step is to select appropriate input and output signals for the analysis. One possibility is to use a small-signal current source connected in parallel at a particular circuit node as an input signal and consider the voltage at that node as the output signal (Fig. 1(a)). In this case, the *HTF(s)* matrix will be an impedance matrix. Equivalently, we can also use a small-signal voltage source connected in series at a circuit branch as the input signal. In this case, the current flowing through that branch represents the output signal (Fig. 1(b)) and the *HTF(s)* matrix will be an admittance matrix.

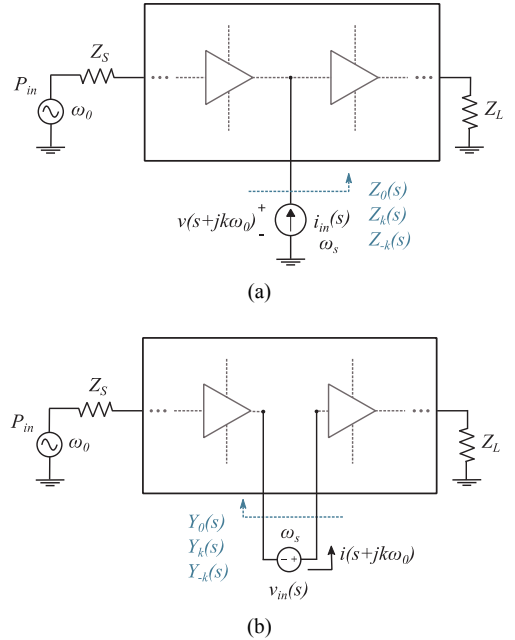

Fig. 1. Illustration of the strategy for obtaining frequency responses required for the construction of the *HTF(s)* matrix. (a) Introduction of a small-signal current source in parallel at a given node of the circuit. (b) Introduction of a small-signal voltage source in series at a given branch of the circuit.

It is important to remark that the *HTF(s)* matrix corresponds to the linearized system without the inclusion of the stabilization parameter. The whole procedure assumes that instability has been previously detected either through a conventional SISO pole-zero identification or using a MIMO approach, like in [15], which detects suitable nodes or branches for stabilization.

Once the input/output representation has been selected, the next step of the methodology is to obtain the components of the central column of the *HTF(s)*:

$$\boldsymbol{H}(s) = [\ldots, H_{-1}(s), H_0(s), H_1(s), \ldots]. \quad (10)$$

To obtain the frequency response of the linearized system around the periodic steady state forced by the large-signal at $f_0$, a large-signal/small-signal harmonic-balance analysis is used, sweeping the frequency of the small-signal current or voltage source at $f_s$. This simulation can be carried out in a commercial simulator as *Advanced Design System* (ADS) [10]. For the case in Fig. 1(a), where the input signal is a current source connected at a particular node, the $2NH+1$ impedance frequency responses in (11) can be obtained in a single simulation run (note that an equivalent formulation can be developed for the case of Fig. 1(b) with admittance frequency responses):

$$H_0(j\omega) = Z_0(j\omega) = \frac{v(j\omega)}{i_{in}(j\omega)}$$
$$H_k(j\omega) = Z_k(j\omega) = \frac{v(jk\omega_0 + j\omega)}{i_{in}(j\omega)} \quad (11)$$
$$H_{-k}(j\omega) = Z_{-k}(j\omega) = \left[\frac{v(jk\omega_0 - j\omega)}{i_{in}(j\omega)}\right]^*$$

where $k = [1,\ldots, NH]$ and $NH$ is the number of harmonics of $f_0$ considered in the harmonic balance simulation.

An *HTF(s)* matrix of $(2NH+1) \times (2NH+1)$ dimension can be built from the frequency responses in (11). To improve efficiency in the calculations that follow, the *HTF(s)* can be truncated further. In practice, not all the $H_k$ and $H_{-k}$ elements have the same relevance in the matrix and some may be negligible. This will depend on the particular case under study, and may differ from one example to another.

From the analysis of multiple examples, a heuristic criterion is proposed here to reduce the order of the *HTF(s)* avoiding the undesired effects associated to the truncation. For every $k$ ($k = 1, 2, \ldots, NH/2$) the magnitude curves of the $H_k$ and $H_{-k}$ frequency responses are compared with the magnitude of $H_0$ on the $[0, kf_0]$ frequency band. The *HTF(s)* can be truncated to the order $(2n+1) \times (2n+1)$ if condition (12) is satisfied for $\forall k$ fulfilling $NH/2 \geq k > n$.

$$|H_k| < |H_0| \text{ and } |H_{-k}| < |H_0| \text{ for all } f_s \in [0, kf_0] \quad (12)$$

As a result of checking the conditions in (12) for $NH/2 \geq k > n$, a reduced vector of frequency responses $\boldsymbol{H}(j\omega)$ is created:

$$\boldsymbol{H}(j\omega) = [H_{-n}(j\omega), \ldots, H_0(j\omega), \ldots, H_n(j\omega)]. \quad (13)$$

It is important to remark that this order reduction is based on empirical data and has proven to work properly in a variety of examples analyzed.

Next, the frequency-domain identification technique can be used to obtain a vector of transfer functions (14) from the vector of frequency responses in (13). Vector Fitting (VF) [25]-[27] will be used here as the identification algorithm. It has the benefit of allowing the identification of a vector of frequency responses with a same set of poles, which makes it especially well-suited for the modeling of MIMO linear systems. Besides, its use of the partial fraction representation makes it numerically better conditioned than other identification algorithms when dealing with large bandwidths and high frequencies, as will be the case here.

$$\boldsymbol{H}(s) = [H_{-n}(s), \ldots, H_0(s), \ldots, H_n(s)] \quad (14)$$

For the identification process, VF assumes that the transfer functions have Hermitian symmetry. Although this is true for $H_0(j\omega)$, the $H_k(j\omega)$ and $H_{-k}(j\omega)$ frequency responses in (13) are non-Hermitian. Consequently, $\boldsymbol{H}(s)$ cannot be directly identified from the vector of frequency responses $\boldsymbol{H}(j\omega)$. To solve this drawback, a Hermitian vector of frequency responses $\tilde{\boldsymbol{H}}(j\omega)$ is constructed from the simulated frequency responses $\boldsymbol{H}(j\omega)$ as follows:

$$\tilde{\boldsymbol{H}}(j\omega) = [\tilde{H}_{-n}(j\omega), \ldots, H_0(j\omega), \ldots, \tilde{H}_n(j\omega)] \quad (15)$$

where,

$$\tilde{H}_k(j\omega) = \frac{1}{2}(H_k(j\omega) + H_{-k}(j\omega))$$
$$\tilde{H}_{-k}(j\omega) = \frac{j}{2}(H_k(j\omega) - H_{-k}(j\omega)). \quad (16)$$

Now, $\tilde{\boldsymbol{H}}(j\omega)$ can be identified with VF to obtain a vector of Hermitian transfer functions $\tilde{\boldsymbol{H}}(s)$. Next, the transformation to Hermitian frequency responses carried out in (16) must be reversed to obtain the original $\boldsymbol{H}(s)$ (14). This is achieved easily by rearranging the state-space matrices resulting from the VF identification. Once the vector of transfer functions $\boldsymbol{H}(s)$ has been determined, the rest of columns of the *HTF(s)* are generated by reordering and frequency shifting the elements of $\boldsymbol{H}(s)$ by $\pm jk\omega_0$, according to (9).

A software routine has been coded in *Matlab* [28] to automatically construct the truncated *HTF(s)* matrix starting from the original set of $2NH+1$ frequency responses simulated in ADS (11). The software routine includes all the steps of the previously described process:

1 – The procedure for the order reduction of the vector $H(j\omega)$.

2 – The Hermitian transformation in (16).

3 – The VF identification to obtain $\tilde{H}(s)$.

4 – The reversal of the Hermitian transformation to obtain $H(s)$.

5 – The matrix arrangements to construct the truncated $HTF(s)$.

At this point it is important to remark that, if we were only interested in the stability analysis of a particular steady-state, we could directly perform a MIMO identification of a vector of multiple frequency responses in (11) without applying the previous processing routine (Hermitian transformations, matrix arrangements, etc.). This is because, if we only wish to determine the stability or instability of our design, we only need to obtain and analyze the poles of the system. From a theoretical point of view, the system poles are common for all the individual transfer functions derived from the frequency responses in (11). Therefore, applying a SISO identification to any of the $H_k(j\omega)$ frequency responses or a MIMO identification to a vector of $H_k(j\omega)$ elements, should provide the same stability results. In practice, losses of controllability or observability may produce pole-zero cancellations that affect some of the individual $H_k(j\omega)$ elements. As a consequence, in a general case, MIMO identification will be more robust and reliable than SISO identification for the detection of instabilities.

## III. Feedback of the $HTF(s)$

For stabilization of dc or small-signal instabilities, standard linear control theory techniques can be applied to a SISO transfer function that relates a small-signal input signal $U(s)$ at $f_s$ to the small-signal output signal $Y(s)$ also at $f_s$ [12]. That is, from a single transfer function one can determine whether the inclusion of a negative feedback to the system provides a stable closed-loop system.

However, the direct root-locus tracing described in [12] is not applicable for predicting the effects of the stabilization networks represented by a negative feedback in a PLTV system (Fig. 2).

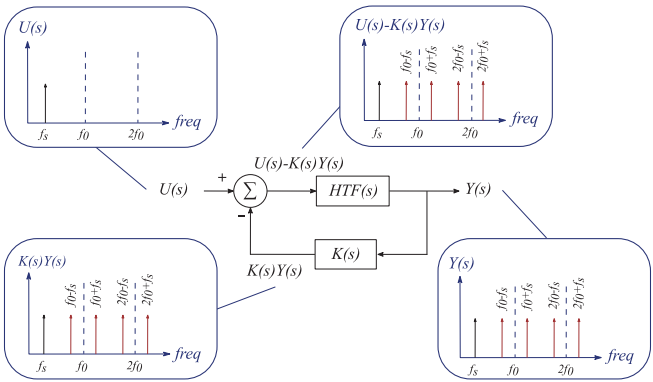

Fig. 2. Block diagram of a PLTV feedback system, where $K(s)$ is the transfer function matrix associated to the stabilization network.

A valid alternative to the root-locus analysis consists of analyzing the poles of the closed-loop system in Fig. 2, for different stabilization networks $K(s)$. In this way, the pole evolution of the closed-loop system versus variations of a stabilization parameter can be predicted.

As shown in Fig. 2, the MIMO nature of the calculus must be respected when applying a negative feedback $K(s)$ to the system defined by the $HTF(s)$ matrix [24], [29]. That is, in theory, the infinite sidebands must be taken into account when calculating the closed-loop system. The closed-loop representation for the large-signal system in Fig. 2 is given by the $HTF_{cl}(s)$ closed-loop transfer function matrix that relates the small-signal input signal $U(s)$ and the output signal $Y(s)$ as follows:

$$HTF_{cl}(s) = \frac{Y(s)}{U(s)} = \left[I + HTF(s)K(s)\right]^{-1} HTF(s) \quad (17)$$

where the feedback transfer function matrix $K(s)$ can, in principle, affect each frequency sideband differently.

Since the inclusion of stabilization resistors is a common procedure to reduce the risk of oscillations, we will particularize (17) for the case of the connection of a stabilization resistor $R_{stab}$ in parallel at a circuit node (Fig. 3) or in series at a circuit branch (Fig. 4). These two cases correspond to a proportional control action with a constant diagonal feedback matrix $K(s)$:

$$K(s) = \begin{bmatrix} \ddots & \vdots & \vdots & \vdots & \iddots \\ \cdots & K & 0 & 0 & \cdots \\ \cdots & 0 & K & 0 & \cdots \\ \cdots & 0 & 0 & K & \cdots \\ \iddots & \vdots & \vdots & \vdots & \ddots \end{bmatrix}. \quad (18)$$

Let us first analyze the case of the parallel connection of a stabilization resistor $R_{stab}$ at a given node of a circuit (Fig. 3). Generalizing the dc or small-signal stabilization techniques discussed in [12], the $HTF(s)$ term in (17) is an impedance matrix whose input is the small-signal current source connected at the node and the output is the voltage at that node. The $K$ constant gain in (18) will be $K = 1/R_{stab}$.

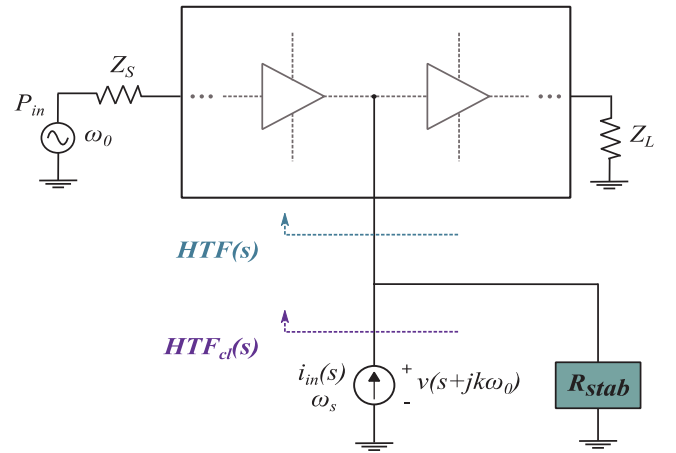

Fig. 3. Parallel connection of a stabilization resistor $R_{stab}$ when $HTF(s)$ is an impedance matrix.

Alternatively, if we consider the case of a stabilization resistor $R_{stab}$ connected in series at a given circuit branch (Fig. 4), the *HTF(s)* term in (17) will be given by an admittance matrix with the small-signal voltage source as the input signal and the current flowing through the branch as the output signal. In this case, the *K* constant gain in (18) will be $K = R_{stab}$.

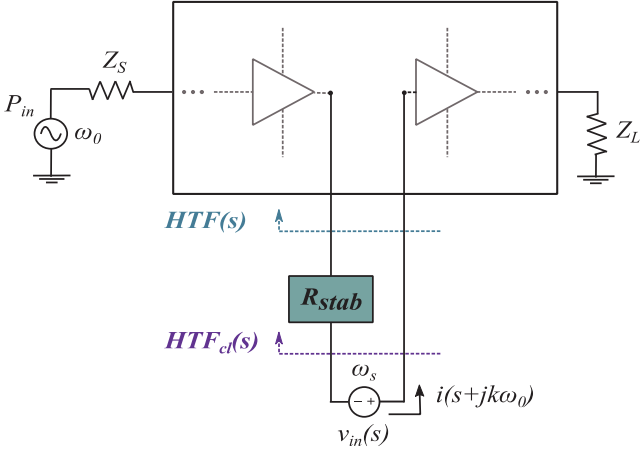

Fig. 4. Series connection of a stabilization resistor $R_{stab}$ when *HTF(s)* is an admittance matrix.

It is important to remind that the *HTF(s)* matrix obtained following the process described in section II.B represents the unstable system before the stabilization parameter is introduced. The closed-loop system, $HTF_{cl}(s)$, with the negative feedback effect of the stabilization parameter is represented by (17). The equation in (17) is a MIMO linear closed loop feedback system in matrix form. It generalizes the small-signal SISO analysis described by equation (6) in [12] to a PLTV system. Once *HTF(s)* and *K(s)* are known, (17) can be solved to obtain the pole evolution of the closed-loop system $HTF_{cl}(s)$ versus variations of the stabilization resistor $R_{stab}$. This is done by a specific *Matlab* [28] function that solves the eigenvalues about the critical frequency band to reduce computation time.

It is important to insist that the solution is completely accurate only if the variations of *K(s)* do not modify the PLTV system represented by the *HTF(s)*. That is to say, the variation of the stabilization resistor $R_{stab}$ does not change the periodic large-signal steady state. If this is not accomplished, the result of solving (17) can only be considered as an approximation. As stated in the introduction, there are typical cases for which the variations of the stabilization parameter have little or no effect on the large-signal steady state under study. One case corresponds to the inter-branch resistors introduced to stabilize odd-mode instabilities in amplifiers with power-combining topologies. A second case concerns the low-frequency instabilities associated to a gain expansion versus input power that are often stabilized introducing stabilization resistors in series with capacitances in the bias paths. These two cases will be addressed in the following section by means of two application examples.

## IV. APPLICATION EXAMPLES

In this section, the method for the calculation of the stabilization parameters described previously is illustrated with two examples. The first example is a low-frequency oscillation at 130 MHz triggered by the input power at $f_0$ = 1.2 GHz of a three-stage amplifier that is stabilized via the introduction of a resistor in series with a bypass capacitor in the gate bias path. The second example consists of an odd-mode parametric oscillation of a power amplifier prototype designed with two stages connected in parallel that is stabilized via the introduction of a stabilization resistor connected in parallel between the gates of the transistors. In neither of the two examples the inclusion of the stabilization resistor affects significantly the steady state regime.

It is important to remark that these two examples are presented here to illustrate the efficiency in the calculation of the stabilization resistor. The examples are particular and their goal is not to extract general conclusions on stabilization strategies.

### A. Low frequency instability

The first example consists of a hybrid L-band three-stage amplifier built in microstrip technology based on GaAs FET transistors. A photograph of the fabricated prototype is shown in Fig. 5. This prototype exhibits a low frequency oscillation at 130 MHz that is originated in the second stage. Depending on the amplifier bias condition, the oscillation can appear just by biasing the amplifier [30] or it can be triggered by the input power [15]. Actually, for nominal bias condition, the oscillation at 130 MHz shows up for $P_{in}$ > - 9 dBm at $f_0$ = 1.2 GHz. We take advantage of this instability that is function of the input power to illustrate the proposed methodology. The unstable output spectrum for $P_{in}$ = - 0.8 dBm is depicted in Fig. 6(a), showing the mixing products around $f_0$. The oscillation is correctly predicted in simulation using the conventional pole-zero stability analysis [6]. Fig. 7 shows the pole-zero map corresponding to the amplifier working with an input power of - 0.8 dBm. A pair of complex conjugate poles with positive real part is obtained at 123 MHz along with the periodic repetitions of these poles with the fundamental frequency.

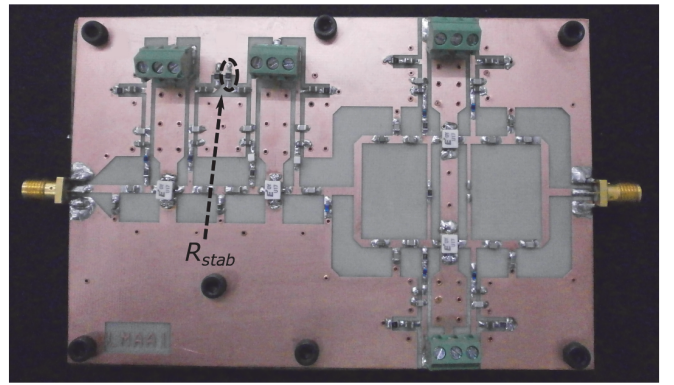

Fig. 5. Photograph of the three-stage amplifier. Stabilization resistor $R_{stab}$ in series with gate decoupling capacitor is outlined.

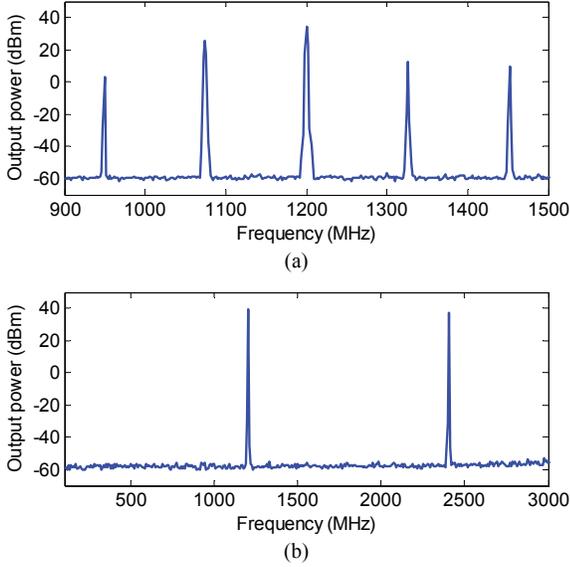

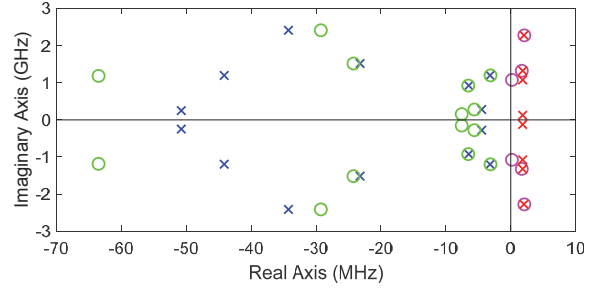

Fig. 7. Pole-zero map of the amplifier biased a nominal conditions and for an input power of $P_{in}$ = - 0.8 dBm at $f_0$ = 1.2 GHz.

Fig. 6. Measured output spectrums showing (a) mixing terms around $f_0$ = 1.2 GHz for $P_{in}$ = - 0.8 dBm due to the parametric low frequency oscillation at 130 MHz. (b) Stable operation for $P_{in}$ = - 0.8 dBm with $R_{stab}$ = 5.1 Ω.

In this case, a resistor $R_{stab}$ connected in series with the second stage decoupling capacitor (Fig. 5) is chosen as the stabilization network. This corresponds to the closed-loop configuration in Fig. 4 with a proportional feedback of $K = R_{stab}$. It should be noted that given that $R_{stab}$ is connected in series with a decoupling capacitor on the bias path, its variations will have little effect on the steady state, as it will be confirmed later. Thus, in order to construct the *HTF(s)* matrix, a small-signal voltage source is introduced in series with the capacitor. Then, a single large-signal/small-signal harmonic balance simulation is performed in ADS to obtain the $\boldsymbol{H}(j\omega)$ required to build the *HTF(s)*. The harmonic balance simulation has been performed with 8 harmonics ($NH = 8$). Construction of the *HTF(s)* is done automatically by the software routine described in section II. In this particular case, condition (12) is satisfied for every $k \geq 1$, as can be observed in Fig. 8 where the magnitudes of all the components of $\boldsymbol{H}(j\omega)$ have been plotted versus frequency. This is explained by the filtering effect of the bias circuitry at higher frequencies. As a result, the *HTF(s)* has been truncated to a $1 \times 1$ matrix, with $H_0$ being the sole component. Once the *HTF(s)* has been constructed, the pole evolution versus $R_{stab}$ can be obtained solving (17). The evolution of the closed-loop poles is plotted with crosses (×) in Fig. 9, where we can observe that a small resistor of a few tenths of ohms is enough to stabilize the circuit. It should be noted that in this case, solving (17) is indeed equivalent to computing the root-locus of just the $H_0$ frequency response.

In principle, the PLTV system resulting from linearizing the periodic large-signal steady state has a MIMO nature due to the sideband conversions as shown in Fig. 2. However, in this example, given that the magnitudes of all the $H_k$ frequency responses are much lower than the magnitude of $H_0$ (Fig. 8), their effect in the feedback of the linearized system is negligible. As a result, the pole evolution can be obtained considering the feedback of a SISO transfer function, $H_0$. Under these circumstances, solving (17) is indeed equivalent to computing the root-locus of the $H_0$ frequency response, as in the small-signal case [12].

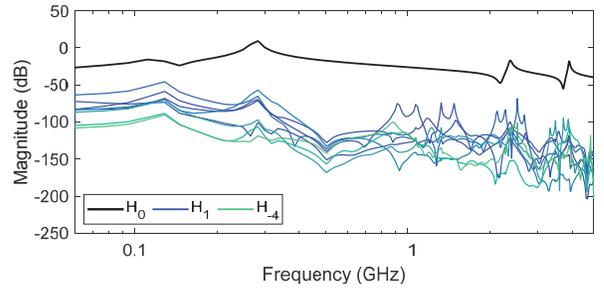

Fig. 8. The magnitude curves of the $H_k$ and $H_{-k}$ frequency responses for $k =1, 2,…, 4$ with respect to the magnitude of $H_0$, all obtained with a voltage source in series, show that condition (12) is met for $k \geq 1$.

In order to validate this result, a complete parametric stability analysis varying $R_{stab}$ has also been performed. For that, large-signal/small-signal harmonic balance simulations, sweeping the parameter $R_{stab}$ have been carried out in ADS. The obtained frequency responses are then identified as in a conventional stability analysis. Results of this parametric stability analysis are superimposed with squares (□) in Fig. 9. A perfect agreement is obtained.

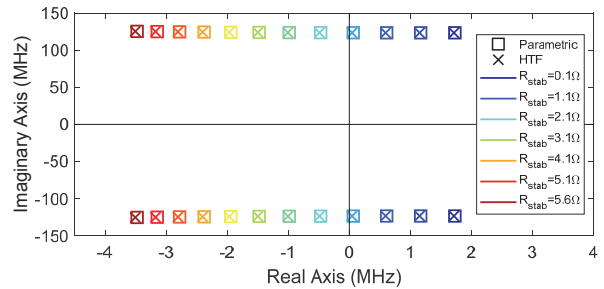

Fig. 9. Pole diagram containing the poles of the closed-loop system derived from the feedback analysis of the *HTF(s)* matrix for variations of the $R_{stab}$ resistor introduced in series (×) and the poles resulting from a conventional parametric stability analysis varying the $R_{stab}$ resistor (□) for $P_{in}$ = - 0.8 dBm.

This perfect match also serves to confirm that the steady-state regime is not modified by the variation of $R_{stab}$. For further confirmation, Fig. 10 shows the simulated voltage spectrum at the input of the second stage transistor for different values of $R_{stab}$. As expected, no differences are appreciable in the spectral components.

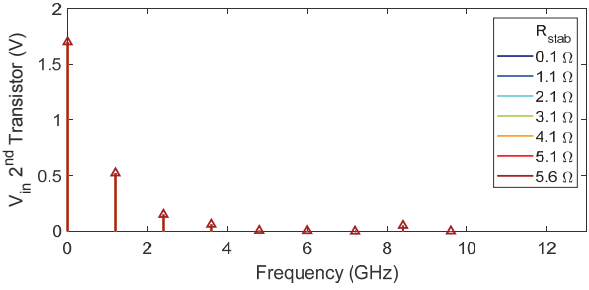

Fig. 10. Simulated spectrum of the large-signal voltage obtained at the input node of the second stage transistor for different values of $R_{stab}$.

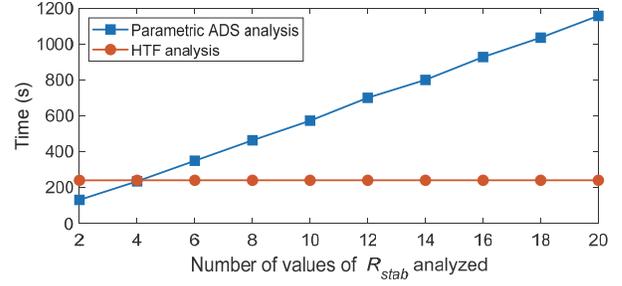

Fig. 11. Run time versus number of values of the $R_{stab}$ analyzed for the conventional parametric analysis of frequency responses (blue squares) and the HTF feedback analysis (red circles) for the three-stage amplifier at $P_{in} = -0.8$ dBm.

Eventually, a small resistor of $R_{stab} = 5.1$ Ω has been introduced in the prototype in order to stabilize it. The resulting stable output spectrum is plotted in Fig. 6(b).

Once the stabilization has been validated, let us analyze the efficiency of the computation. For the HTF method, we sum up the computation time required to simulate the $2NH+1$ frequency responses $H_k(j\omega)$ in ADS, the construction of the $HTF(s)$ in *Matlab* and the calculation of the poles of the feedback system (17) also in *Matlab*. For the conventional stability analysis, the run time of the parametric ADS simulations sweeping the stabilization resistor, plus the corresponding pole-zero identification with *Matlab*, are taken into account. For the later ADS simulation, we consider $NH = 8$ harmonics, and 101 frequency points in the following frequency band [100 MHz, 600 MHz]. In summary, all the measurable processes of both methods have been taken into account in order to objectively compare the computation times. The computer used to calculate the different run times has an Intel i5-7200U processor with a processing speed of 2.5 GHz and 16 GB of RAM memory.

The computation time versus the number of $R_{stab}$ values is plotted in Fig. 11 for the two methods. For simplicity we call them *HTF feedback* and *conventional parametric*. In this particular example, the large-signal/small-signal harmonic balance simulations are very time consuming. The computation time of the HTF feedback method remains almost constant and independent of the amount of parameter values analyzed. This is so because the construction of $HTF(s)$ and the calculation of the poles of the feedback system are almost immediate given the reduced size of the $HTF(s)$ matrix. The only time consuming process in this case is the simulation in ADS of the frequency responses required for the construction of the $HTF(s)$ matrix. This simulation is only carried out once and is independent of the amount of parameter values analyzed, obviously. On the contrary, the conventional parametric stability analysis depends linearly with the number of parameter values. As Fig. 11 clearly shows, the conventional parametric method is faster when a small amount of parameter values is to be analyzed, but the HTF feedback method becomes more efficient as the number of values increases. For twenty values of the stabilization resistor, the conventional parametric stability analysis takes around 20 minutes while the HTF feedback only requires approximately 4 minutes.

### B. Odd-mode parametric oscillation

A specific prototype has been designed and implemented in hybrid microstrip technology to illustrate the proposed methodology. The prototype contains two GaN HEMT transistors (CHJ40010) connected in parallel and biased in class C. A photograph of the fabricated prototype is shown in Fig. 12. Whilst this cell is stable in dc for the nominal bias conditions, it exhibits an odd-mode parametric oscillation, very close to $f_0/2$, when the input power $P_{in}$ at $f_0 = 1$ GHz is higher than -3 dBm (Fig. 13(a)). Fig. 14 shows the pole-zero map corresponding to the amplifier cell working with an input power of -2 dBm. Complex conjugate poles with positive real part are obtained at 497 MHz together with its Floquet repetitions.

A $R_{stab}$ resistor connected between the gates of the transistors (Fig. 12) is chosen as the stabilization network, since it is the optimum solution for stabilization of odd-mode oscillations without affecting the even mode large-signal steady state. This corresponds to the closed-loop configuration in Fig. 3 with a proportional feedback of $K = 1/R_{stab}$. To obtain the frequency responses required to construct the $HTF(s)$ matrix, a current source is introduced between the gates of the two transistors in place of $R_{stab}$. Next, the impedance frequency responses required in this case to construct the $HTF(s)$ matrix are obtained. The harmonic balance simulation has also been performed with 8 harmonics ($NH = 8$).

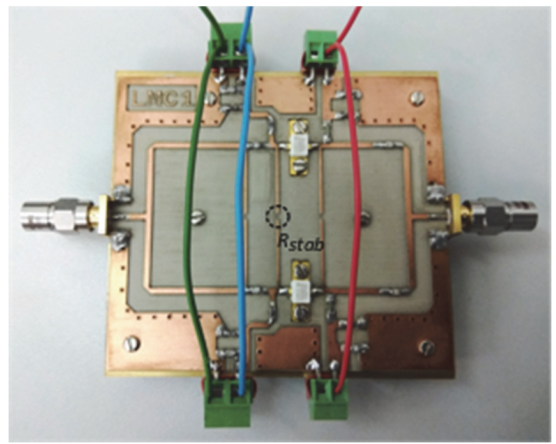

Fig. 12. Photograph of the fabricated prototype signaling the stabilization resistor $R_{stab}$.

9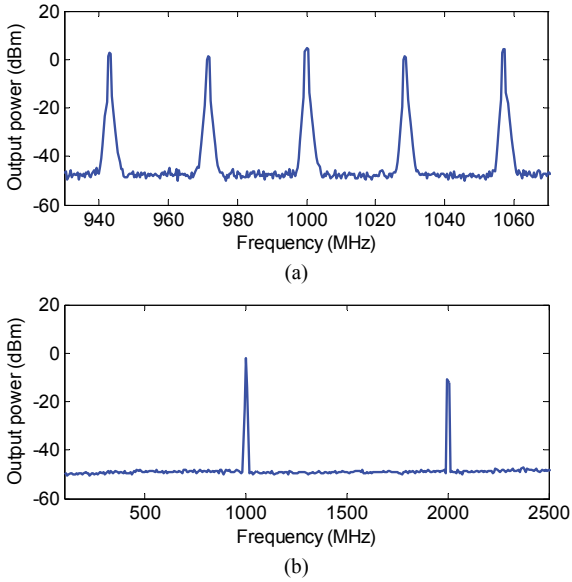

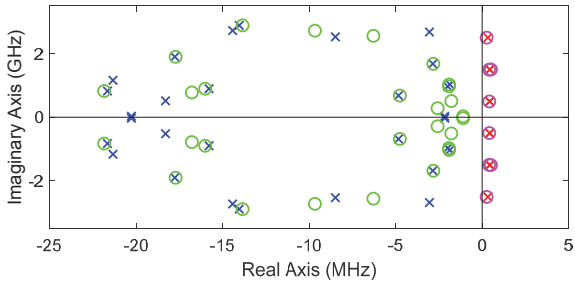

Fig. 13. Measured output spectrums showing (a) oscillation at $f_0/2$ for $P_{in}$ = -2 dBm. (b) Stable operation for $P_{in}$ = -2 dBm with $R_{stab}$ = 510 Ω.

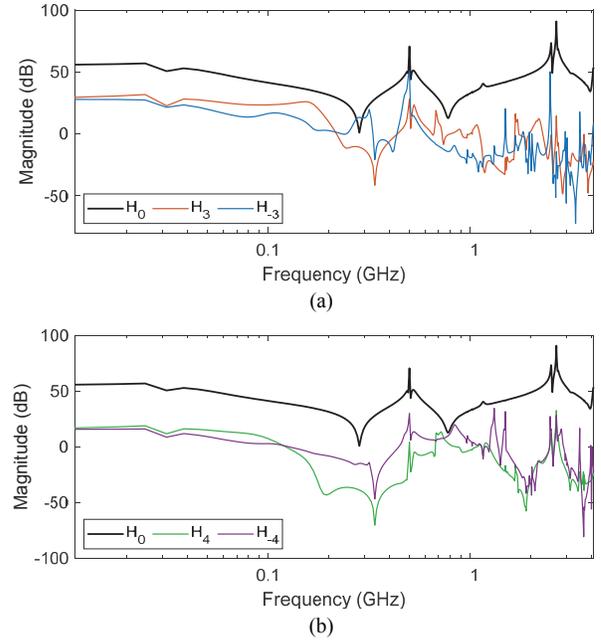

Fig. 14. Pole-zero map resulting from the frequency domain identification of a frequency response obtained for $P_{in}$ = -2 dBm.

In order to determine the size of the truncated *HTF(s)* matrix required for a correct stabilization prediction, the magnitude curves of the frequency responses $H_k$ and $H_{-k}$ for $k = 1,…, 4$ are compared with the magnitude curve of $H_0$. In this particular case, as can be observed in Fig. 15(a) the magnitudes of the $H_k$ and $H_{-k}$ for $k ≤ 3$ are larger than the magnitude of the $H_0$. However, for $k = 4$ the magnitudes of the $H_k$ and $H_{-k}$ fulfill condition (12) (Fig. 15(b)). Therefore, it is determined that an *HTF(s)* matrix of size 7 × 7 should be sufficient to accurately analyze the large-signal stabilization of this odd-mode instability.

Indeed, as can verified in Fig. 16, the poles identified with a conventional parametric pole-zero analysis of the $H_0$ frequency response simulated in ADS (□) and the closed-loop poles obtained by applying a proportional feedback of $K = 1/R_{stab}$ to the *HTF(s)* matrix of size 7 × 7 (×) match perfectly. If the feedback analysis is carried out solely parting from the $H_0$ frequency response, the obtained results will not match those obtained with the conventional parametric pole-zero analysis, as shown in Fig. 17. Furthermore, the obtained results would erroneously indicate that adding a resistor between the gate nodes of the amplifiers cannot stabilize the circuit, since unstable poles remain on the right-hand side of the *s* plane for any value of $R_{stab}$. As a consequence, a simple root locus analysis applied to $H_0$ is not appropriate for this example. Eventually, a resistor of $R_{stab}$ = 510 Ω has been introduced in the prototype in order to stabilize it. The resulting stable output spectrum is plotted in Fig. 13(b).

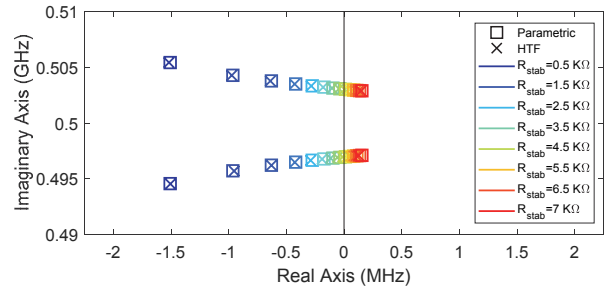

Fig. 15. (a) The magnitude curves of the $H_3$ and $H_{-3}$ frequency responses with respect to the magnitude of $H_0$, all obtained with a current source in parallel, show that condition (12) is not met for $k = 3$. (b) The magnitude curves of the $H_4$ and $H_{-4}$ frequency responses with respect to the magnitude of $H_0$, all obtained with a current source in parallel, show that condition (12) is met for $k ≥ 4$.

Fig. 16. Pole diagram containing the poles of the closed-loop system derived from the feedback analysis of the *HTF(s)* matrix for variations of the $R_{stab}$ resistor introduced in parallel (×) and the poles resulting from a conventional parametric stability analysis varying the resistor (□) for $P_{in}$ = -2 dBm.

As in the previous example, the computation time required by the two methods is compared as a function of the number of parameter values for this particular case. For the conventional parametric stability analysis, the run time of the total simulation versus the number of $R_{stab}$ values considered in the analysis is plotted with blue squares in Fig. 18. In this case, the small-signal/large-signal harmonic-balance simulation in ADS is performed from 0.1 MHz to 0.6 GHz, with 101 frequency points. In turn, the total computation time required by the HTF feedback method with an *HTF(s)* matrix of size 7 × 7 is plotted with red circles in Fig. 18. The same qualitative behavior is obtained again. For parameter sweeps with few values, the conventional method is faster. On the contrary, the benefits of the HTF feedback method become appreciable when the number of parameter values in the sweep is high.

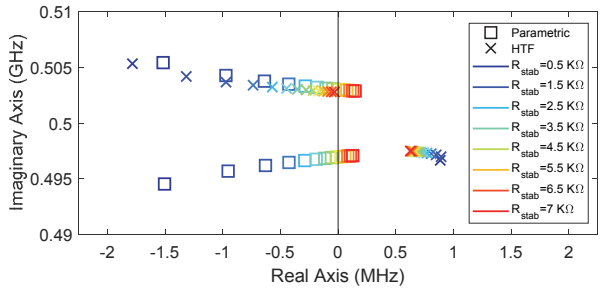

Fig. 17. Pole diagram containing the poles of the closed-loop system derived from the feedback analysis of the $HTF(s)$ matrix, comprised solely of the $H_0(s)$ transfer function, for variations of the $R_{stab}$ resistor introduced in parallel (×) and the poles resulting from a conventional parametric stability analysis varying the resistor (□) for $P_{in}$ = -2 dBm. The stabilization resistor $R_{stab}$ required cannot be correctly determined in this case.

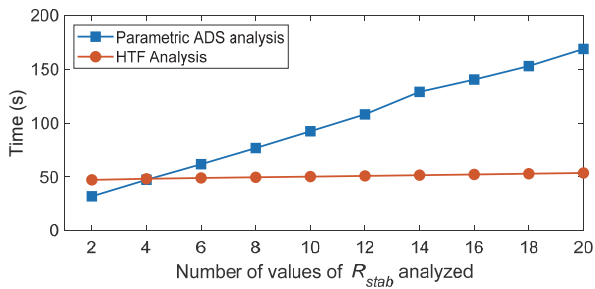

Fig. 18. Run time versus number of values of the $R_{stab}$ analyzed for the conventional parametric stability analysis (blue squares) and the HTF feedback analysis (red circles) for the two-stage amplifier at $P_{in}$ = -2 dBm.

## V. DISCUSSION

The proposed methodology for the calculation of the stabilization parameter values is not intended for all types of stabilization strategies and particular cases. As it has been stressed along the paper, a key requirement of the procedure is that the steady state under analysis must not be modified by the stabilization parameter. Two examples of these situations have been presented in this work. However, there are other instances where the steady state changes only slightly with the inclusion stabilization elements. This happens for instance in the parallel RC stabilization networks inserted in series with transistor gates or bases to reduce gain at low frequencies while having low effect at the operating frequency and harmonics [31]. In this case, the proposed method can be used as an efficient way to obtain approximate values for the stabilization elements.

The amount of gain in computation time with respect to the full ADS simulation is not general either and depends on many factors such as the complexity of the circuit, the non-linear models, the convergence difficulties in harmonic balance or the number of parameter values. This has been evidenced in the two examples presented here. Actually, the $HTF(s)$ feedback analysis is not intended for those cases where the full ADS simulation is not significantly time consuming. However, when ADS simulation run times are large, the proposed method provides a considerable gain in time for the calculation of the stabilization parameters. It is important to note that, the heuristic condition (12) has been set very strictly in order to guarantee a pole calculation with high accuracy, as shown in Fig. 9 and Fig. 16. This condition can be relaxed to reduce further the order of the $HTF(s)$. In this way, we will have a gain in computing time at the cost of losing some accuracy. This compromise can be acceptable in most cases. In addition, from the evaluation of (12) we can detect those cases where the $HTF(s)$ can be reduced to $H_0$ as in the example of section IV.A. In those situations, calculation of the stabilization parameter is virtually immediate, as in a conventional root locus tracing of a SISO linear system.

The fitting of the transmission lines may be another source of concern in rational identification. Generally, for a simple stability analysis this is not particularly challenging. As demonstrated in [32], even in circuits with elements described with transcendental functions, local instability is always governed by a finite number of right-half-plane (RHP) poles. The detection of those unstable poles using rational functions is very effective, especially if the poles are resonant with low damping factors, which is the common case in realistic microwave amplifiers. However, in our case we need to obtain a complete model of the linearized system (the $HTF(s)$) previous to the stabilization calculation. From a fitting point of view, this is more challenging that just detecting the presence of resonant poles. If the frequency band of interest is very wide and in the presence of transmission lines, the number of stable poles required to obtain an accurate rational model can be very high. For this reason, the use of Vector Fitting [25]-[27] is key to obtaining an accurate model of the $HTF$ because the algorithm is particularly well suited to handle high-order transfer functions at high frequencies.

Finally, it is important to remind that a common source of uncertainty in the pole-zero stability analysis comes from the loss of controllability and observability at some particular observation nodes that translates to pole-zero cancellations in the transfer function. A MIMO strategy that tries to minimize this problem and to determine the most suitable locations for stabilization was presented in [15]. This strategy can be applied in combination with the techniques presented here. We also need to be very conscious at the time of computing the frequency response because most of the times the accuracy of a pole-zero stability analysis depends more on the correctness of the simulated frequency response than on the pole-zero identification process itself. Inaccurate non-linear models, imperfect circuit descriptions or numerical errors in the simulated frequency response (convergence errors, truncation noise, etc.) can lead to incorrect pole-zero plots with false conclusions about the stability of the system.

## VI. CONCLUSION

A novel and efficient approach for the calculation of the stabilization parameters required to eliminate spurious oscillations that appear on the periodic large-signal regime has been proposed. The methodology has been illustrated through the stabilization of a three-stage amplifier with low-frequency instability and a second two-stage amplifier with an odd-mode parametric oscillation for increasing input power. The two examples have served to illustrate the capability and efficiency of the proposed approach.

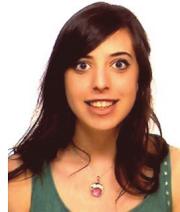

**Libe Mori** received the Ph.D. degree in electronics from the University of the Basque Country (UPV/EHU), relating to linear and nonlinear stability analysis of microwave power amplifiers. She is currently with Mondragon Unibertsitatea working as a lecturer and researcher. Her main research interests include analysis and design of microwave circuits.

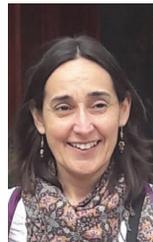

**Ibone Lizarraga** received the M.Sc. degree in physics with electronics and Ph.D. degree in control engineering from the University of the Basque Country (UPV/EHU), Bilbao, Spain, in 1994 and 2001, respectively. She is currently with the Electricity and Electronics Department, UPV/EHU, where she has been an Associate Professor since 1998. Her research interests include the analysis and control of mechatronic system, stability analysis and linear and nonlinear system identification.

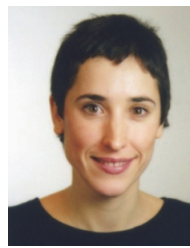

**Aitziber Anakabe** received the Ph.D. degree in electronics from the University of the Basque Country (UPV/EHU), Bilbao, Spain, in 2004. In 1999, she joined the Electricity and Electronics Department, UPV/EHU, where she was involved with the stability analysis of nonlinear microwave circuits. In 2004, she joined the French Space Agency (CNES), Toulouse, France, as a Post-Doctoral Researcher. In 2005, she rejoined the Electricity and Electronics Department, UPV/EHU, where, since 2005, she has been an Associate Professor. Her research deals with nonlinear analysis and modeling of microwave circuits and measurement techniques.

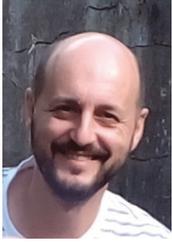

**Juan-Mari Collantes** received the Ph.D. degree in electronics from the University of Limoges, France, in 1996. Since February 1996, he has been an Associate Professor with the Electricity and Electronics Department, University of the Basque Country (UPV/EHU), Bilbao, Spain. In 1996 and 1998 he was an Invited Researcher with Agilent Technologies (formerly the Hewlett-Packard Company), Santa Rosa, CA. In 2003, he was with the French Space Agency (CNES), Toulouse, France, where he was involved with power amplifier analysis, simulation, and modeling. His areas of interest include nonlinear analysis and design of microwave circuits, microwave measurement techniques and noise characterization.

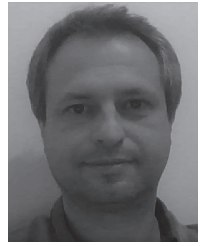

**Vincent Armengaud** received his Ph. D. in Electronics from the University of Limoges (France) in 2008. He has first worked for Thales Alenia Space France on MMIC design. In 2012 he joined the French Space Agency (CNES), Toulouse, France. His main research interests are the low noise amplifiers and medium level circuits.

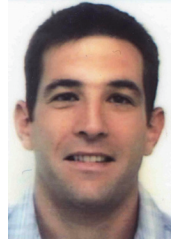

**Geoffroy Soubercaze-Pun** received his Ph.D. in Electronics from the University of Toulouse (France) in 2007. He has first worked for Thales Alenia Space France on frequency synthesis. In 2008 he joined the French Space Agency (CNES), Toulouse, France. His main research interests are power amplifiers and non linear analysis of microwave circuits.